\documentclass[aps,prl,preprint,superscriptaddress]{revtex4}
\usepackage{graphicx}
\usepackage{color}
\usepackage{amsfonts}

\begin{document}
\newpage
\title{Sub-60 mV/decade switching with a cold metal as the injection source}
\author{Fei Liu}
\email{feiliu@pku.edu.cn}
\affiliation{Institute of Microelectronics and Key Laboratory of Microelectronics Devices and Circuits (MoE), Peking University, Beijing 100871, China}
\begin{abstract}
 Power dissipation is a great challenge for the continuous scaling down and performance improvement of CMOS technology, due to thermionic current switching limit of conventional MOSFETs. In this work, we show that this problem can be overcome by using cold metals as the transistor's injection source, which are different from conventional metals and can filter high energy electrons to break the Boltzmann tyranny. It is proved that the subthreshold swing (SS) of thermionic current of transistor using cold metal contact can be extremely smaller than 60 mV/decade at room temperature. Specifically, two-dimensional (2D) transition metal chalcogenide (TMD) cold metals of NbX$_2$ and TaX$_2$(X=S, Se, Te) are proposed as the injection source of FETs. Quantum transport simulations indicate that promising switching efficiency and on-state current can be achieved using TMD cold metal injection source, which is beneficial for energy efficient applications.  \\
\\
KEYWORDS: Cold metal, energy efficient, subthreshold swing, quantum transport simulations, two dimensional materials
\end{abstract}
\newpage
\maketitle
\section{INTRODUCTION}

The power consumption becomes increasingly severe with the improvement of integration degree, and becomes an important bottleneck for the advancement of CMOS technology along Moore's law\cite{ACSeabaugh,AMIonescu}. High power consumption will not only increase chip temperature and failure rate, but also increase chip design, packaging and cooling costs. Especially with the rise of various mobile electronic devices, such as smart phones, wearable devices, internet of things, etc. The demand for reducing chip power consumption is increasingly strong. Both the static and dynamic power consumption of the chip is related to the supply voltage, so lowering the supply voltage is an effective way to reduce the power consumption of the chip\cite{ACSeabaugh,AMIonescu}. However, this is subject to the transistor subthreshold swing (SS) limit, which cannot be lower than 60 mV/decade at room temperature.

Steep slope devices have attracted much attention for designing power-constrained applications by using tunneling\cite{JAppenzeller,DSarkar}, impact ionization\cite{KGopalakrishnan} and negative capacitance\cite{SSalahuddin}. Recently, cold source FETs (CS-FETs) are proposed to obtain the sub-thermionic switching by source density of states (DOS) engineering cutting-off the Boltzmann tail of the current\cite{FLiu02}. Such cold source can be realized by using emerging Dirac materials\cite{FLiu01,CQiu}, properly doped semiconductors\cite{FLiu02,FLiu01} or a narrow-energy conduction band of dangling bonds\cite{DLogoteta}. However, these proposed cold source requires properly doping and heterogenous integration. An ideal solution is to find a kind of cold metals to replace conventional metals as the contact of transistors. Such cold metals would filter high energy electrons in the subthreshold region.

\begin{figure}[t!]
\centering
\includegraphics[width=5in]{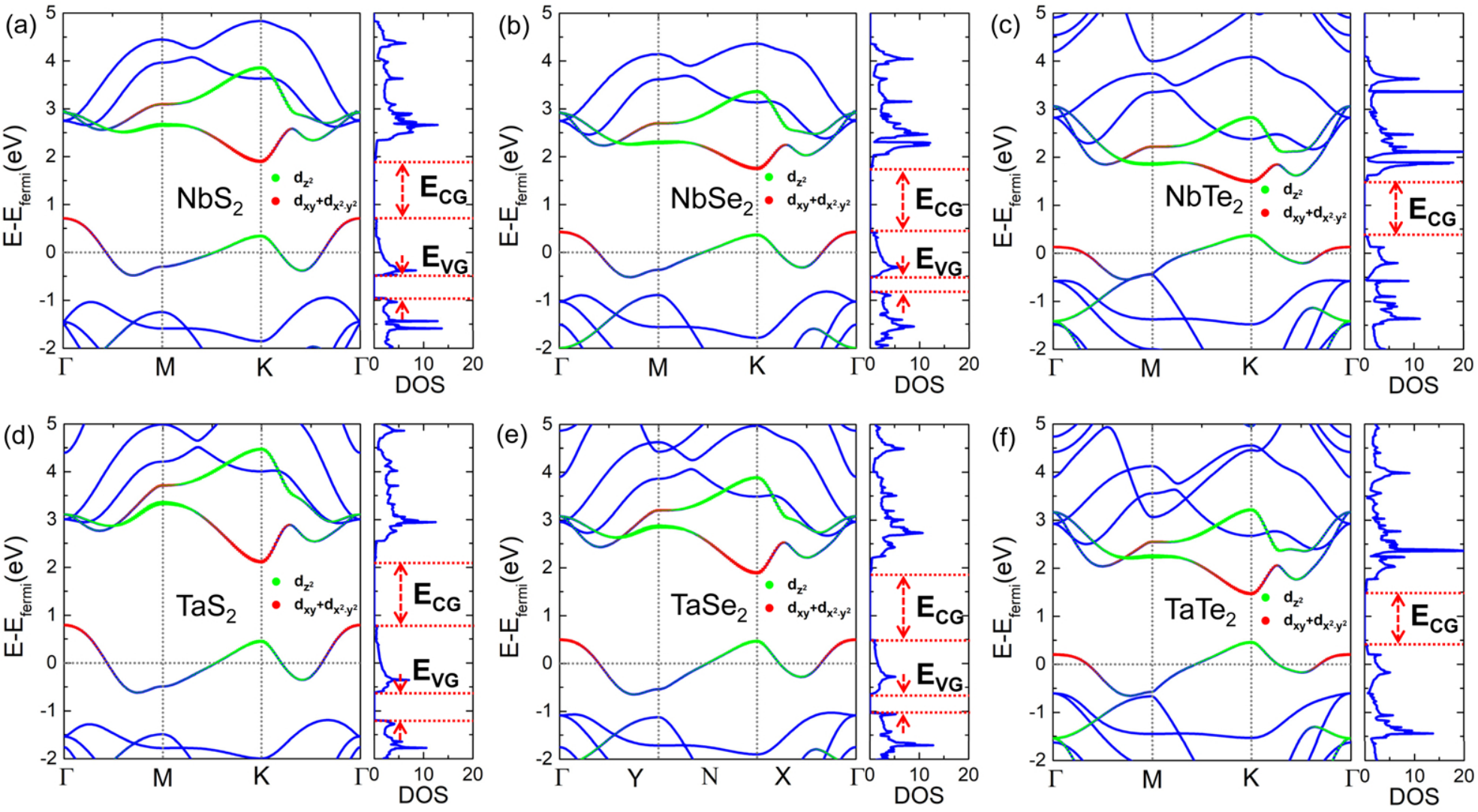}
\caption{Band structures for monolayer metallic MX$_2$ (M = Nb, Ta; X = S, Se, Te). Fermi energy is set to zero. Different from conventional metals, there is an energy gap in the conduction band (E$_{CG}$)or valence band (E$_{VG}$). Therefore, electron in the gap can be filtered as these metallic materials are applied as the transistor's injection source.}\label{FIG03}
\end{figure}

In this work, an emerging steep slope device using cold metal contact is proposed. Here, cold metals are applied to replace conventional metals as the FET contact. Different from conventional metals, cold metals have an energy gap around the Fermi level and work like p-type or n-type doped semiconductors. Therefore, electron in the energy region can be effectively filtered, which results in sub-60 mV/decade switching. As the emergence of 2D materials, such metallic materials with required properties can be found in monolayer (ML) NbX$_2$ and TaX$_2$ (X = S, Se, Te), which be achieved by mechanical exfoliation\cite{KSNovoselov,XXi}, chemical exfoliation\cite{JNColeman, RJSmith,JWu} or chemical vapour deposition method (CVD)\cite{JShi,JZhou,HWang,MHossain,YZhang}. We investigated 2D heterojunction CS-FETs using cold metal contact of metallic transition metal dichalcogenides (TMDs) as the injection source by quantum transport simulations using the non-equilibrium Greens function (NEGF) formalism. It is demonstrated that CS-FETs with TMD cold metal contacts have promising device performance of energy efficient switching and high on-state current. Our work reveals a novel design rule for future steep slope electronic devices using cold metal contact.

\section{RESULTS and DISCUSSION}
\textbf{Cold metals}
Metals widely exist in nature and are a very important substance for the semiconductor industry. In MOSFETs, metallic materials are applied as the contacts, gates and interconnects. Usually, these bulk metals have excellent electrical conductivity with continuous DOS around the Fermi level. Recently, 2D metallic materials have been intensively investigated due to their exotic properties\cite{ZPZhang}. When reducing the dimension from bulk metallic TMDs to atomic thin limit, 2H-NbSe$_2$ has been found to exhibit intriguing quantum phenomena of superconductivity and charge density wave\cite{XXi, MMUgeda, XiXX, YCao}. Monolayer 2H-NbS$_2$ has been synthesized epitaxially and applied as the injection source of 2D FETs using the lateral and vertical NbS$_2$-WS$_2$ heterostructures\cite{YZhang}.

\begin{figure}[t!]
\centering
\includegraphics[width=5in]{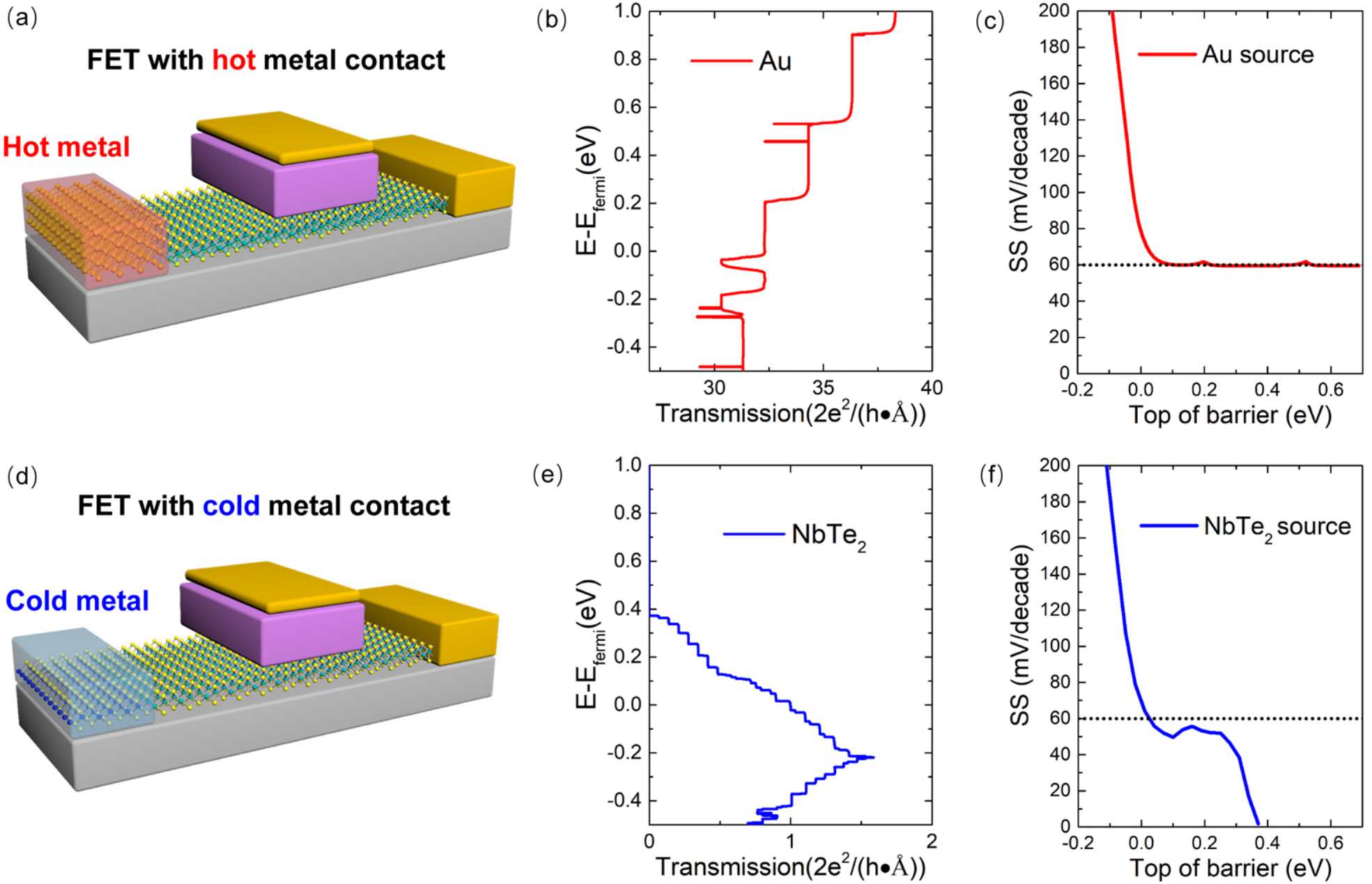}
\caption{Comparison between conventional hot metal and cold metal contacts for transistors. Schematic illustration of transistors with (a) hot and (d) cold metals. Conventional metal of Au have a continuous transport channel around the Fermi level (b), while cold metal of NbTe$_2$ have an energy region without channel above the Fermi level(e). Subthreshold swing of transistor using Au is always larger than 60 mV/decade (c), while SS using NbTe$_2$ as the injection source can break the thermionic limit (f).}\label{FIG02}
\end{figure}
Here, we first investigated electronic properties of metallic TMDs by the density functional theory (DFT) calculations, which were performed using the Vienna ab initio simulation package (VASP)\cite{GKresse01}.  The core-valence interaction was described by the projector augmented wave (PAW) method\cite{PEBlochl,GKresse02}. Exchange-correlation functional of metallic TMDs was described by the generalized gradient approximation of Perdew-Burke-Ernzerhof (GGA-PBE)\cite{GGA01,GGA02}. While, the band structure calculations of semiconducting TMDs were performed by using the Heyd-Scuseria-Ernzerhof (HSE06) hybrid functional\cite{HSE01,HSE02}. Energy cut off for plane-wave expansion was set to 500 eV. Monkhorst-Pack (MP) k point meshes were used for Brillouin zone sampling with a grid of 15$\times$15$\times$1 for both structure relaxations and self-consistent calculations. Atomic structures were relaxed  until the calculated residual forces are smaller than 0.01 eV/\AA. A vacuum layer larger than 15 \AA\ is used to avoid adjacent image interaction.

As one kind of TMDs, 2H-NbX$_2$ (X = S, Se, Te) has a common chemical formula of MX$_2$ with a transition-metal layer atoms sandwiched between two layers of chalcogen X atoms with covalent bonds.
It is well known that 2H-MoX$_2$ are semiconducting with the lowest subbands contributed by the d$_{z^2}$ , d$_{xy}$, and d$_{x^2-y^2}$ orbitals of transition-metal atoms. There are 5 $d$ electrons in the 4-th shell and the valence bands are filled. However, Nb atoms have one less $d$ electron in the outermost shell and the Fermi level is in the  valence bands. 2H-NbX$_2$ are metallic and the band structures of 2H-NbX$_2$ have similar shapes as those of MoX$_2$ as shown in Fig. 1. The single band around the Fermi level is contributed by the hybridization of  the d$_{z^2}$ , d$_{xy}$, and d$_{x^2-y^2}$ orbitals. Similarly, 2H-TaX$_2$ are also  metallic and have similar band structures with the Fermi level across the topmost valence band. Unlike conventional metals with continuous density of states around the Fermi level, there is an energy gap above the chemical potential of these materials as shown in Fig. 1. At the same time, a valence band gap appears below the Fermi level of NbS$_2$, NbSe$_2$, TaS$_2$ and TaSe$_2$. Therefore, these metallic materials are naturally p-type or n-type semiconductor without artificial doping. When these metallic materials are applied as the injection source of FETs, electrons with energy in the gap can be filtered.

Next, we compared the difference between 2H-NbX$_2$ and traditional metal as the FET contact. Fig. 2 (a) and (d) shows the schematic device structures with Au and 2H-NbTe$_2$ as the injection source, respectively. The current through the device can be calculated by the Landauer-Buttiker formula\cite{SDatta}:
\begin{equation}\label{EQ01}
  I_D = \frac{2q}{h} \int T(E)[f_S-f_D] dE
\end{equation}
Where T(E) is the transmission coefficient, f$_S$ and f$_D$ are the Fermi functions of the source and drain, respectively.  Fig. 2(b) plots the transmission as a function of energy of 3 nm golden film, which is calculated by using NEGF-DFT implemented in Nanodcal\cite{NEGFDFT}. As expected, golden film have a continuous transport channel around the Fermi level of E$_F$ = 0 eV.  In contrast, Fig. 2(e) shows that there is no transport channel above due to the energy gap in the conduction band of 2H-NbTe$_2$. The SS is given by \cite{ACSeabaugh,AMIonescu}:
\begin{equation}\label{EQ01}
  SS = \frac{\partial V_G}{\partial log10(I_D)}
\end{equation}
where V$_G$ is the gate voltage. In the case of neglecting the tunneling current, the SS of thermionic current as a function of the top of channel barrier is calculated using the obtained transmission of Au contact in Fig. 2(c). Here, the top of barrier is modulated by the gate voltage. It can be found that the SS can not break the switching limit of 60 mV/decade using Au contact. Differently, 2H-NbTe$_2$ has an energy gap in the conduction band (E$_{CG}$). When the top of barrier is in the E$_{CG}$, the current does not change as the lowering of channel barrier and SS is infinitely large. While, as the top of barrier gets smaller than E = 0.37 eV, the current is suddenly increased and SS is extremely small as shown in Fig. 2 (f). It is also found that the SS keeps smaller than 60 mV/decade in the energy region of due to the linearly increased channels, which is beneficial for the energy efficient switching.  The physics behind this sub-60 mV/decade using 2H-NbTe$_2$ is that the high energy electron above the Fermi level is filtered by the energy gap. Hence, the off-state current does not change when the top of barrier is in the energy gap region. 2H-NbTe$_2$ as the FET injection source is different from conventional metal and works like a cold source by filtering high energy electrons to realize sub-60 mV/decade switching.
\begin{figure}[t!]
\centering
\includegraphics[width=5in]{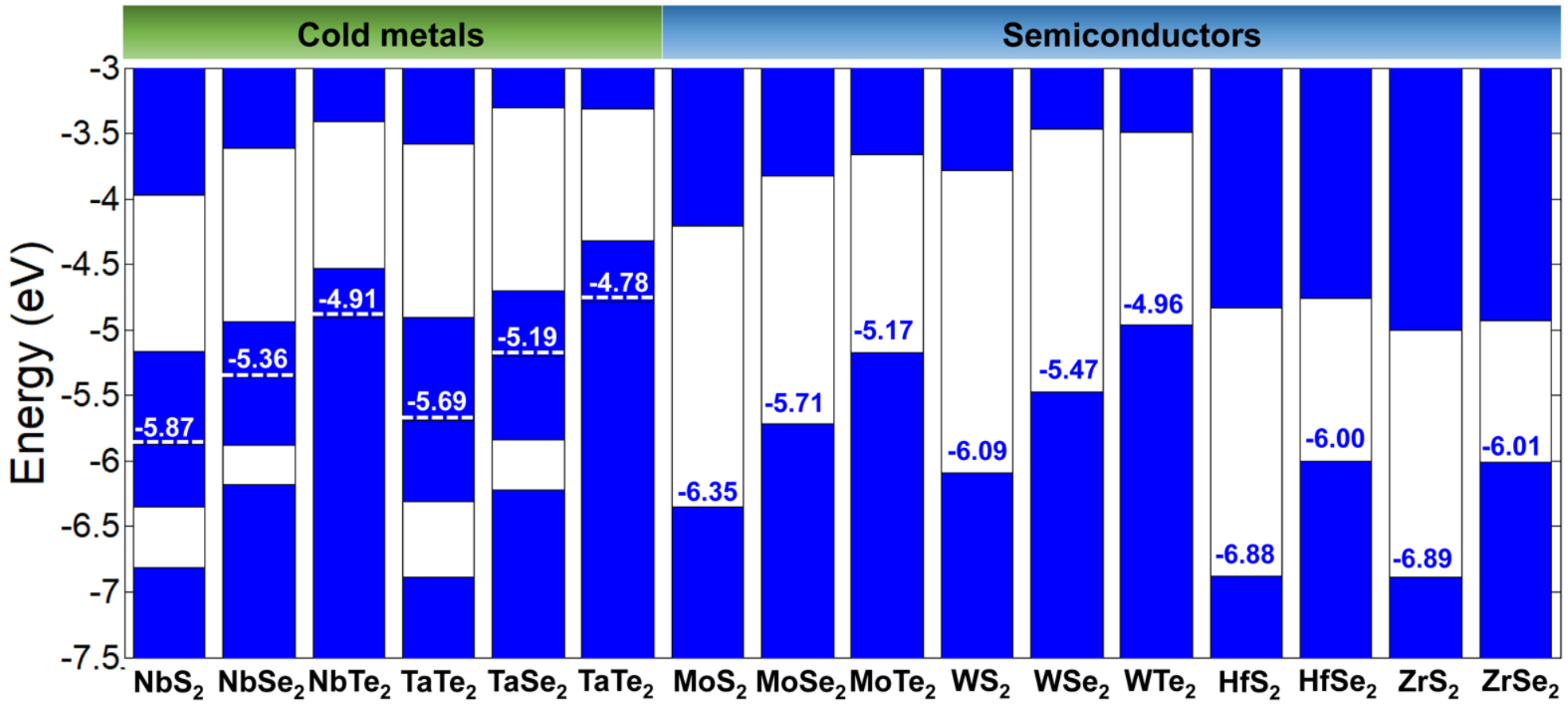}
\caption{Calculated band alignment for metallic and semiconducting TMDs.  Schottky and Ohmic contacts can be realized by properly choosing metallic and semiconducting TMDs. }\label{FIG03}
\end{figure}

 Band offsets of metals and semiconductors are essential for device design. Therefore, we studied band alignments of metallic and semiconducting TMDs by the first principles calculations using VASP as shown in Fig. 3. Here, NbX$_2$, TaX$_2$, MoX$_2$ and WX$_2$ are in 2H-phase while HfX$_2$ and  ZrX$_2$ are in 1T phase\cite{ZhangC}. As the atomic number of X increases (from S to Te), the Fermi level of NbX$_2$ increases. The Fermi levels of TaX$_2$ are higher than those of NbX$_2$ due to much higher energy of the 5d orbital of the transition metal. It can be found that the contact between metallic and semiconducting TMDs is a Schottky type for the same X, such as NbS$_2$-MoS$_2$ and NbTe$_2$-MoTe$_2$.  NbS$_2$ has the lowest fermi level and is much easier to realize p-type Ohmic contact, such as NbS$_2$-MoSe$_2$ and NbS$_2$-WS$_2$. In contrast, TaTe$_2$ has the highest chemical potential and can be applied as n-type Ohmic contact, such as TaTe$_2$-ZrS$_2$ and TaTe$_2$-HfS$_2$.

\textbf{Steep slope FETs with cold metal contact} To assess the role of cold metals of metallic TMDs as the injection source, we constructed hetero-junction FETs using TMDs. It has been shown that metallic NbX$_2$ can be exfoliated from bulk counterpart\cite{KSNovoselov,XXi} or synthesized by a chemical vapor deposition (CVD) method\cite{YZhang}. Moreover, both lateral and vertical metal-semiconductor NbS$_2$-WS$_2$ hetero-structures are achieved\cite{YZhang}, which paves a promising way to construct CS-FETs using 2D metal-semiconductor TMD hetero-junctions. An important metric for choosing proper cold metal for n-type FET is the energy difference between the Fermi level and band gap in conduction band, which is 0.71 eV, 0.42 eV and 0.37 eV for 2H-NbS$_2$, 2H-NbSe$_2$ and 2H-NbTe$_2$, respectively. The energy difference can not be too large in order to filter those electrons in sub-threshold region. So, 2H-NbTe$_2$ is applied as the injection source and MoSe$_2$ is used as the channel material of CS-FETs as shown in Fig. 4(a). The two materials have relatively small lattice mismatching. According to the band alignments, the contact between NbTe$_2$ and MoSe$_2$ is a Schottky type and the n-type Schottky barrier is 1.08 eV as shown in Fig. 3. To reduce the Schottky barrier, 5\% compressive strain and 5\% tensile strain are applied on NbTe$_2$ and MoSe$_2$, respectively. As a result, the n-type Schottky barrier is reduced to 0.09 eV.

\begin{figure*}[t!]
\centering
\includegraphics[width=6in]{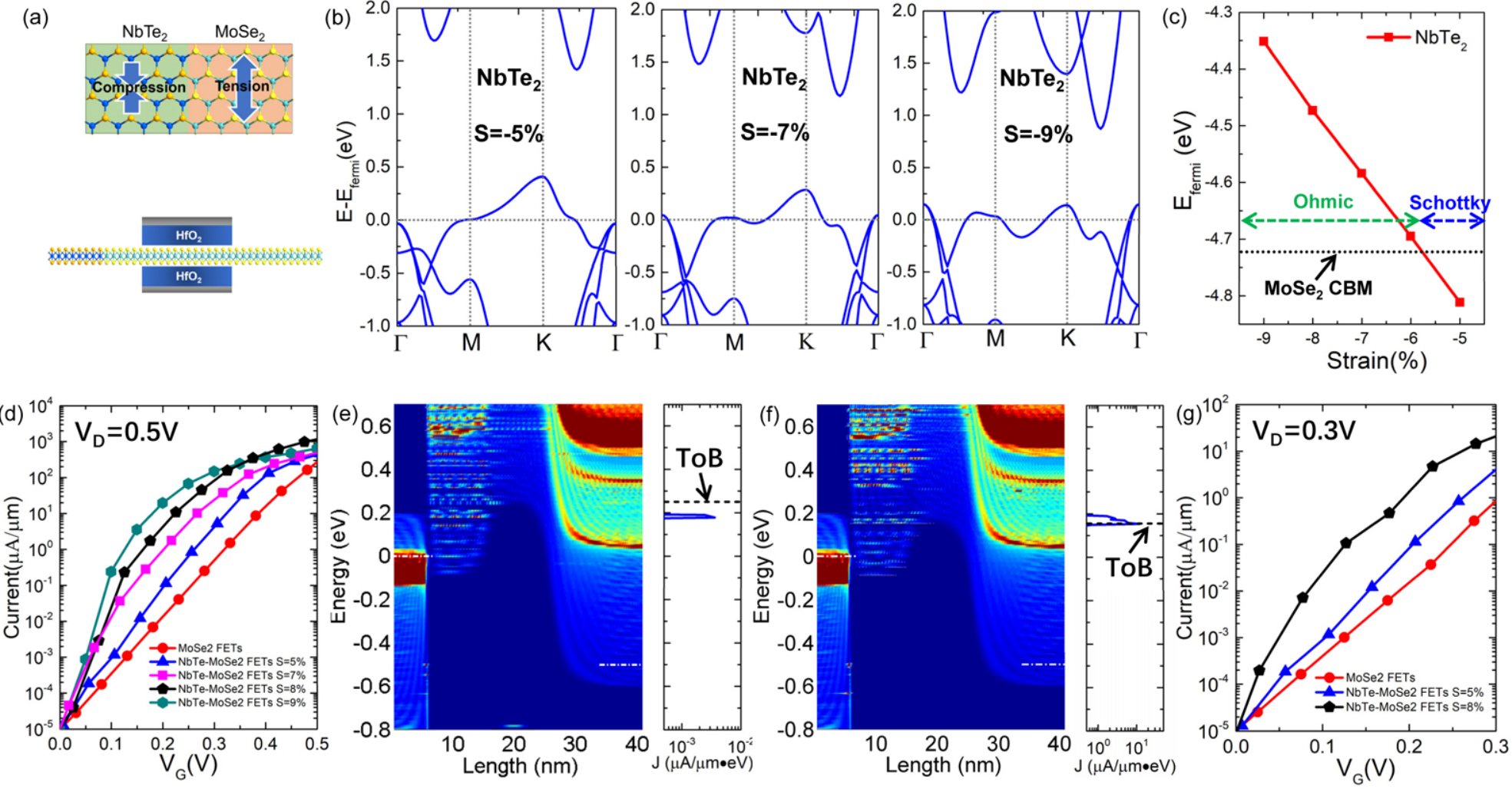}
\caption{Structure and performance of NbTe$_2$-MoSe$_2$ CS-FETs. (a)Schematic device structure of hetero-junction CS-FETs using NbTe$_2$ and MoSe$_2$.   (b) Band structures of NbTe$_2$ under compressive strain. (c) The Fermi energy of NbTe$_2$ as a function of strain. Under compressive strain, the contact type between NbTe$_2$ and MoSe$_2$ changes from Schottky contact to Ohmic contact. (d) I$_D$-V$_G$ of NbTe$_2$-MoSe$_2$ CS-FETs with NbTe$_2$ under compressive strain at V$_D$ =0.5 V. Device performance can be improved by the strain. Local density of states of NbTe$_2$-MoSe$_2$ CS-FETs with -8\% strain on NbTe$_2$  with top of barrier (ToB) (e) above (V$_G$ = -0.03 V) and (f) below (V$_G$ = -0.13 V)source valence band maximum (VBM). (g) I$_D$-V$_G$ of NbTe$_2$-MoSe$_2$ CS-FETs with NbTe$_2$ at V$_D$ =0.3 V.}\label{FIG03}
\end{figure*}

The transfer characteristics (I$_D$-V$_G$) of CS-FETs and MoSe$_2$ FETs are compared at V$_D$ = 0.5 V in Fig. 4(d). High dielectric material of 3 nm HfO$_2$ is applied as the top and bottom oxides.  The source, drain and gate lengths of simulated FETs are 15 nm, 15 nm and  10 nm, respectively. Metallic NbTe$_2$ is 4 nm in the source of NbTe$_2$-MoSe$_2$ CS-FETs. Three band tight binding model is applied to describe 2H-NbTe$_2$ and 2H-MoSe$_2$\cite{GLiu}, and MoSe$_2$ in the source and drain regions is doped to n-type. Device performance calculations are performed by self-consistently solving the Poisson equation and Schrodinger equation within the NEGF formalism\cite{FLiu03}. The Off-current is fixed to 10 pA/$\mu$m at V$_G$ = 0 V. Compared with MoSe$_2$ FETs, device performance is improved in NbTe$_2$-MoS$_2$ CS-FETs. On-current of CS-FETs reaches 4.4$\times$10$^2$ $\mu A /\mu m$ larger than 3.4 $\times$10$^2$ $\mu A /\mu m$ MoSe$_2$ -FETs. The SS of CS-FETs is about 51 mV/decade between 0.0 V $<$ V$_G$ $<$ 0.1 V, while the SS of MoSe$_2$ FETs is around 65 mV/decade. Even though the SS of CS-FETs breaks the switching limit, it is much larger than theoretical prediction in Fig. 3.  The reason is that the band gap edge in conduction band of 2H-NbTe$_2$ under -5\% strain is 0.41 eV and electrons for deep subthreshold region are filtered rather than those for subthreshold region with current larger than 10$^{-5}$ $\mu A /\mu m$. Therefore, the VBM of 2H-NbTe$_2$ has to be reduced in order to improve the device performance of NbTe$_2$-MoS$_2$ CS-FETs.

Fig. 4(b) shows the band structure of NbTe$_2$ under compressive strain. As the compressive strain is increased from 5\% to 9\%, the energy difference between the Fermi level and VBM is decreased from 0.41 eV to 0.15 eV.  The strain also changes the Fermi level of NbTe$_2$ as shown in Fig. 4(c). The Fermi energy increased with the compressive strain, so the contact between NbTe$_2$ and MoSe$_2$ changes from Schottky type at S = -5\% to Ohmic type at S = -9\%. Fig. 4(d) shows that current switching is greatly improved in NbTe$_2$-MoS$_2$ CS-FETs as the increasing of compressive strain on NbTe$_2$. SS is as small as 26 and 23 mV/decade under 8\% and 9\% compressive strain on NbTe$_2$, respectively. Fig. 4 (e) plots the local DOS (LDOS) along the channel at V$_G$ = 0.03 V with -8\% strain on NbTe$_2$. The top of channel barrier is above the VBM of NbTe$_2$ and current is composed of 100\% direct tunneling current from source to drain in Fig. 4(e). As the gate voltage is increased to V$_G$ = 0.13 V, the ToB gets lower than the source VBM in Fig. 4(f). As a result, the current is mainly thermionic current over the channel barrier. I$_{on}$ can be as large as 1.2$\times$10$^3$$\mu$A/$\mu$m, which is larger than 5.3$\times$10$^2$$\mu$A/$\mu$m of the low power requirements of the International Technology Roadmap for Semiconductors (ITRS) 2013\cite{ITRS2013} for L$_g$ = 10 nm at V$_D$ = 0.75 V. Advantages of CS-FET gets more significant as V$_D$ is scaled to 0.3 V as shown in Fig. 4(g). I$_{on}$ reaches 23 $\mu$A/$\mu$m over ten times larger than that of MoSe$_2$ FETs and I$_{on}$/I$_{off}$ ratio is over 10$^6$.
\begin{figure*}[t!]
\centering
\includegraphics[width=5in]{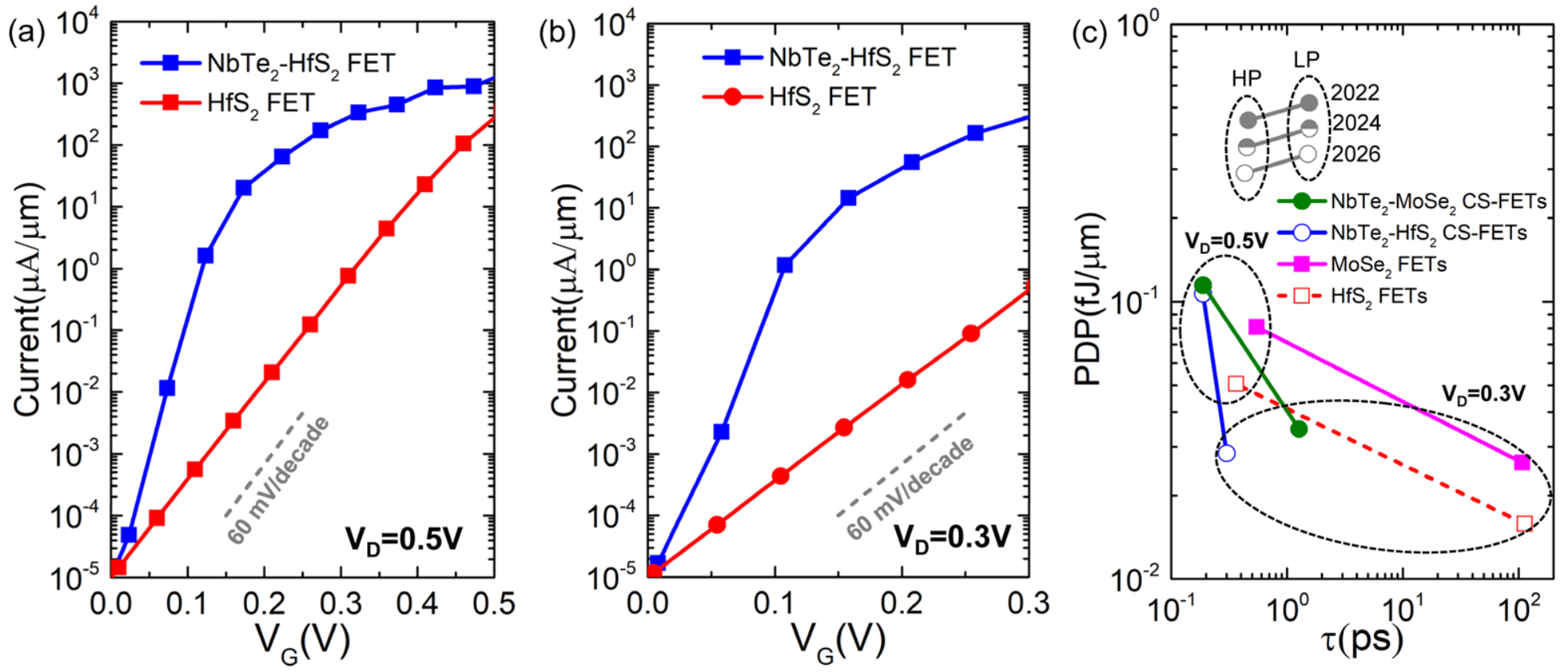}
\caption{Device performance of NbTe$_2$-HfS$_2$ CS-FETs. I$_D$-V$_G$ of NbTe$_2$-HfS$_2$ CS-FETs and HfS$_2$ FETs at (a) V$_G$ = 0.5 V and (b) V$_G$ = 0.3 V. Power-delay product and switching speed of NbTe$_2$-MoSe$_2$ CS-FETs, NbTe$_2$-HfS$_2$ CS-FETs, MoSe$_2$ FETs and HfS$_2$ FETs compared with the requirements of the ITRS 2013\cite{ITRS2013}.  }\label{FIG03}
\end{figure*}

 We also constructed NbTe$_2$-HfS$_2$ CS-FETs because the contact between the two materials is intrinsically Ohmic type as shown in Fig. 3. At the same time, lattice mismatching between ML-NbTe$_2$ and ML-HfS$_2$ is tiny. ML-NbTe$_2$ is under -9\% biaxial strain and ML-HfS$_2$ is applied as the channel material. Both ML-NbTe$_2$ and ML-HfS$_2$ are described by eleven band tight-binding model\cite{TB11} based on maximally localized Wannier functions\cite{NMarzari,AAMostofi}. Due to the Ohmic contact, device performance of NbTe$_2$-HfS$_2$ CS-FETs is promising and much better than that of HfS$_2$ FETs as shown in Fig. 5(a, b).  I$_{on}$ reaches 1.2$\times$10$^3$ $\mu$A/$\mu$m and 3.1$\times$10$^2$ $\mu$A/$\mu$m at V$_D$ = 0.5 V and 0.3 V, respectively. We also investigated the switching speed and energy efficiency of NbTe$_2$-MoSe$_2$ and NbTe$_2$-MoSe$_2$ CS-FETs in Fig. 5(c). The switching speed is measured by the intrinsic delay, defined as $\tau=(Q_{on}-Q_{off})/I_{on}$, where $Q_{on/off}$ is the channel total charge in the on/off state. Energy efficiency  is measured by power-delay product (PDP) given by $PDP=V_{D}(Q_{on}-Q_{off})$. It can be found that switching speeds and switching energies of NbTe$_2$-MoSe$_2$ and NbTe$_2$-HfS$_2$ CS-FETs at L$_G$ =10 nm and V$_D$ = 0.5 V are much better than the requirements of the ITRS 2013\cite{ITRS2013}. For example, the switching speed of NbTe$_2$-HfS$_2$ CS-FETs is 2.4 and 8.3 times faster than the requirement of ITRS 2013 for high-performance (HP) and low-power (LP) applications, respectively. While, the PDP is only 24\% and 21\% for HP and LP applications, respetively.  As the supply voltage is decreased to 0.3 V, PDP and switching speed can be further improved while on-state current is decreased. MoSe$_2$ FETs and HfS$_2$ FETs also have lower PDP but the on-state currents are much smaller than those of CS-FETs, which also results in longer switching speed.

At last, we discussed the possible magnetic state of NbTe$_2$ which may have significant impact on realizing sub-60 mV/decade switching using such cold metal. The transition metals of Nb and Ta have one unpaired outmost electron, so magnetic moment may appear. It has been shown that intrinsic NbX$_2$ has no magnetic moment, while NbS$_2$ and NbSe$_2$ structures can be magnetized under tensile strain\cite{ZhouY}. It is predicted that NbS$_2$ and NbSe$_2$ under 10\% tensile strain are ferromagnetic room temperature. Here, we applied NbTe$_2$ under compressive strain as the contact. The energy difference per unit cell between ferromagnetic and  anti-ferromagnetic NbTe$_2$ under -9\% strain is 30 mV. The Curie temperature T$_C$ under the mean field approximation can be estimated by T$_C$ = 2$\Delta$E/(3k$_B)$ according to the Heisenberg model\cite{ZhouY}, which is 235 K for NbTe$_2$ under -9\% strain. So, NbTe$_2$ under -9\% strain is paramagnetic at room temperature.

\section{CONCLUSION}
In summary, steep-slope transistors using cold metals as the injection source are proposed and investigated by quantum transport simulations. It is revealed that cold metals such as 2D TMDs of NbX$_2$ and TaX$_2$ can effectively cut off the Boltzmann tail of electrons due to the energy gap above the Fermi level. Therefore, sub-60 mV/decade switching can be obtained by using such cold metals as the injection source.  Comprehensive device simulations show that a steep subthreshold swing (23 mV/decade) and high on-state current over 1.0$\times$10$^3$ $\mu$A/$\mu$m is achievable in both NbTe$_2$-MoSe$_2$ and NbTe$_2$-HfS$_2$ CS-FETs, which exhibit promising performance with respect to the requirements of ITRS 2013 for both high-performance and low-power applications. Our findings pave a way for using cold metal materials to design steep slope electronic devices with the rapid development of emerging materials.\\
 \\

\makeatother

\begin{thebibliography}{35}

\bibitem{ACSeabaugh}
A. C. Seabaugh, Q. Zhang, Low voltage tunnel transistors for
beyond CMOS logic. Proc. IEEE \textbf{98}, 2095-2110 (2010).

\bibitem{AMIonescu}
A. M. Ionescu, H. Riel, Tunnel field-effect transistors as energy
efficient electronic switches. Nature \textbf{479}, 329-337 (2011).

\bibitem{JAppenzeller}
J. Appenzeller, Y. M. Lin, J. Knoch, P. Avouris, Band-to-band
tunneling in carbon nanotube field-effect transistors. Phys. Rev. Lett.,
\textbf{93}, 196805 (2004).

\bibitem{DSarkar}
D. Sarkar, X. Xie, W. Liu, W. Cao, J. Kang, Y. Gong, S. Kraemer, P. M. Ajayan, K. Banerjee, K., A subthermionic tunnel field-effect transistor with an atomically thin channel. Nature, \textbf{526}, 91 (2015).

\bibitem{KGopalakrishnan}
K. Gopalakrishnan, P. B. Griffin, and J. D. Plummer, Impact ionization
MOS (I-MOS)-Part I: device and circuit simulations. IEEE Trans. Electron Devices.
\textbf{52}, 69每76 (2005).

\bibitem{SSalahuddin}
S. Salahuddin and S. Datta, Use of negative capacitance to provide
voltage amplification for low power nanoscale devices. Nano Lett.
\textbf{8}, 405-410 (2008).

\bibitem{FLiu02}
F. Liu, C. G. Qiu, Z. Y. Zhang, L. Peng, J. Wang, Z. Wu, H. Guo, First principles simulation of energy efficient switching by source density of states engineering. IEEE International Electron Devices Meeting (IEDM), 763-766 (2018).

\bibitem{FLiu01}
F. Liu, C. Qiu, Z. Zhang, et al., Dirac electrons at the source: breaking the 60-mV/decade switching limit. IEEE Trans. Electron Devices., \textbf{65}, 2736-2743 (2018).

\bibitem{CQiu}
C. Qiu, F. Liu, L. Xu, et al., Dirac source field-effect transistors as energy-efficient and high-performance electronic switches. Science \textbf{361}, 387 (2018).

\bibitem{DLogoteta}
D. Logoteta, M. G. Pala, J. Choukroun, P. Dollfus, G. Iannaccone, A steep-slope MoS$_2$-Nanoribbon MOSFET based on an intrinsic cold-contact effect.  IEEE Electron Device Lett. \textbf{40}, 1550 (2019).

\bibitem{KSNovoselov}
 K. S. Novoselov, D. Jiang, F. Schedin, T. J. Booth, V. V. Khotkevich, S. V. Morozov, A. K. Geim, Two dimensional atomic crystals. Proc. Natl Acad. Sci. USA, \textbf{102} 10451-3 (2005).

\bibitem{XXi}
 X. Xi, Z. Wang, W. Zhao, J .H. Park, K. T. Law, H. Berger, L. Forro, J. Shan, K. F. Mak, Ising pairing in superconducting NbSe$_2$ atomic layers. Nat. Phys., \textbf{12}, 139-143 (2015).

\bibitem{JNColeman}
J. N. Coleman, M. Lotya, A. O'Neill, S. D. Bergin, P. J. King, U. Khan, K. Young, A. Gaucher, S. De, R. J. Smith, et al. Two-Dimensional Nanosheets Produced by Liquid Exfoliation of Layered Materials. Science \textbf{331}, 568每571 (2011).

\bibitem{RJSmith}
R. J. Smith, P. J. King, M. Lotya, C. Wirtz, U. Khan, S. De, A. O'Neill, G. S. Duesberg, J .C. Grunlan, G. Moriarty, et al. Large-scale exfoliation of inorganic layered compounds in aqueous surfactant solutions. Adv. Mater. \textbf{23}, 3944-3948 (2011).

\bibitem{JWu}
J. Wu, J. Peng, Z. Yu, Y. Zhou, Y. Guo, Z. Li, Y. Lin, K. Ruan, C. Wu, Y. Xie, Acid-assisted exfoliation toward metallic sub-nanopore TaS$_2$ monolayer with high volumetric capacitance. J. Am. Chem. Soc.  \textbf{140}, 493-498 (2018).

\bibitem{HWang}
H. Wang, X. Huang, J. Lin, J. Cui, Y. Chen, C. Zhu, F. Liu, Q. Zeng, J. Zhou, P. Yu, X. Wang,  High-quality monolayer superconductor NbSe$_2$ grown by chemical vapour deposition. Nature commun.  \textbf{8}, 394 (2017).

\bibitem{JZhou}
J. Zhou, J. Lin,  X. Huang, Y. Zhou, Y. Chen, J. Xia, H. Wang, Y. Xie, H. Yu, J. Lei, D. Wu,  A library of atomically thin metal chalcogenides. Nature, 556(7701), p.355-359 (2 018).

\bibitem{JShi}
J. Shi, X. Chen, L. Zhao, Y. Gong, M. Hong, Y. Huan, Z. Zhang, P. Yang, Y. Li, Q. Zhang, Q. Zhang, L. Gu, H. Chen, J. Wang, S. Deng, N. Xu, Y. Zhang, Chemical vapor deposition grown wafer-scale 2D tantalum diselenide with robust charge-density-wave order. Adv. Mater. \textbf{30}, 1804616 (2018).

\bibitem{MHossain}
M. Hossain, J. Wu, W. Wen, H. Liu, X. Wang, L. Xie, Chemical vapor deposition of 2D vanadium disulfide and diselenide and Raman characterization of the phase transitions, Adv. Mater. Interfaces, \textbf{5}, 1800528 (2018).

\bibitem{YZhang}
Y. Zhang, L. Yin, J. Chu, T. A. Shifa, J. Xia, F. Wang, Y. Wen, X. Zhan, Z. Wang, J. He,  Edge-epitaxial growth of 2D NbS$_2$-WS$_2$ lateral metal-semiconductor heterostructures. Adv. Mater. \textbf{30}, 1803665 (2018).

\bibitem{ZPZhang}
Z. Zhang, P. Yang, M. Hong, S. Jiang, G. Zhao, J. Shi, Q. Xie, Y. Zhang, Recent progress in the controlled synthesis of 2D metallic transition metal dichalcogenides. Nanotech., \textbf{30}, 182002 (2019).

\bibitem{MMUgeda}
M. M. Ugeda, A. J. Bradley, Y. Zhang, S. Onishi, Y. Chen, W. Ruan, C. Ojeda-Aristizabal, H. Ryu, M. T. Edmonds, H. Z. Tsai, A. Riss, S. K. Mo, D. Lee, A. Zettl, Z. Hussain, Z. X. Shen, M. F. Crommie, Characterization of collective ground states in single-layer NbSe$_2$. Nat. Phys. , \textbf{12}, 92-97 (2016).

\bibitem{XiXX}
X. X. Xi, H. Berger, L. Forro, J. Shan, K. F. Mak, Gate Tuning of Electronic Phase Transitions in Two-Dimensional NbSe$_2$. Phys. Rev. Lett. \textbf{117}, 106801 (2016).

\bibitem{YCao}
Y. Cao, A. Mishchenko, G. L. Yu, E. Khestanova, A. P. Rooney, E. Prestat, A. V. Kretinin, P. Blake,  M. B. Shalom, C. Woods, J. Chapman, G. Balakrishnan, I. V. Grigorieva, K. S. Novoselov, B. A. Piot, M. Potemski, K. Watanabe, T. Taniguchi, S. J. Haigh, A. K. Geim, R. V. Gorbachev, Quality heterostructures from two-dimensional crystals unstable in air by their assembly in inert atmosphere, Nano Lett., \textbf{15}, 4914-4921 (2015).

\bibitem{GKresse01}
G. Kresse, J. Furthm$\ddot{u}$ller, Efficient iterative schemes for ab initio total-energy
calculations using a plane-wave basis set. Phys. Rev. B \textbf{54}, 11169每11186 (1996).

\bibitem{PEBlochl}
P. E. Bl$\ddot{o“}$chl, Projector augmented-wave method. Phys. Rev. B \textbf{50}, 17953每17979
(1994).

\bibitem{GKresse02}
G. Kresse, D. Joubert, From ultrasoft pseudopotentials to the projector augmented-wave method. Phys. Rev. B \textbf{59}, 1758每1775 (1999).

\bibitem{GGA01}
J. P. Perdew, K. Burke, M. Ernzerhof, Generalized gradient approximation made simple. Phy. Rev. Lett. \textbf{77}, 3865每3868 (1996).

\bibitem{GGA02}
S. Grimme, Semiempirical GGA-type density functional constructed with a long-range dispersion correction. J. Comput. Chem. \textbf{27}, 1787每1799 (2006).

\bibitem{HSE01}
J. Heyd, G. E. Scuseria, M. Ernzerhof, Hybrid functionals based on a screened Coulomb potential. J. Chem. Phys. \textbf{118}, 8207每8215 (2003).

\bibitem{HSE02}
J. Heyd, G. E. Scuseria, M. Ernzerhof, Erratum: Hybrid functionals based on a screened Coulomb potential. J. Chem. Phys. \textbf{124}, 219906 (2006).

\bibitem{SDatta}
S. Datta, Quantum Transport: Atom to Transistor. Cambridge, U.K.:
Cambridge Univ. Press (2005).

\bibitem{NEGFDFT}
J. Taylor, H. Guo, J. Wang, Ab initio modeling of quantum transport properties of molecular electronic devices, Phys. Rev. B \textbf{63}, 245407 (2001).

\bibitem{ZhangC}
 C. Zhang, C. Gong, Y. Nie, K. N. Min, C. Liang, Y. J. Oh, H. Zhang.,  W. Wang, S. Hong, L. Colombo, R. M. Wallace., K. Cho, Systematic study of electronic structure and band alignment of monolayer transition metal dichalcogenides in Van der Waals heterostructures, 2D Mater. \textbf{4}, 4, 015026 (2017).


\bibitem{GLiu}
G. Liu, W. Shan, Y. Yao, W. Yao, D. Xiao, A three-band tight-binding model for monolayers of group-VIB transition metal dichalcogenides, Phys. Rev. B,  \textbf{88}, 085433 (2013).

\bibitem{FLiu03}
F. Liu, Y. Wang, X. Liu, J. Wang, H. Guo, Ballistic transport in monolayer black phosphorus transistors, IEEE Trans. Electron Devices, \textbf{61}, 3871-3876 (2014).
2014.2353213.

\bibitem{ITRS2013}
 Process Integration, Devices, and Structures, (2013).
Available: http://www.itrs.net/

\bibitem{TB11}
S. Fang, R. K. Defo, S. N. Shirodkar, S. Lieu, G. A. Tritsaris, and E. Kaxiras, Ab initio tight-binding hamiltonian for transition metal dichalcogenides, Phys. Rev. B \textbf{92}, 205108 (2015).

\bibitem{NMarzari}
 N. Marzari and D. Vanderbilt, Maximally localized generalized Wannier functions for composite energy bands, Phys. Rev. B \textbf{56}, 12847 (1997).

\bibitem{AAMostofi}
A. A. Mostofi, J. R. Yates, G. Pizzi, Y.-S. Lee, I. Souza, D. Vanderbilt, and N.Marzari,Anupdated version of WANNIER90: A tool for obtaining maximally-localised Wannier functions, Comput. Phys. Commun. \textbf{185}, 2309 (2014).

\bibitem{ZhouY}
Y. Zhou, Z. Wang, P. Yang, X. Zu, L. Yang, X. Sun, F. Gao, Tensile strain switched ferromagnetism
in layered NbS$_2$ and NbSe$_2$, ACS Nano, \textbf{6}, 6, 9727 (2012).



\end{thebibliography}

\end{document}